\documentclass[prb,preprint,showpacs,preprintnumbers,amsmath,amssymb]{revtex4}

\usepackage{graphicx} % Include figure files
\usepackage{dcolumn} % Align table columns on decimal point
\usepackage{bm} % bold math

\usepackage{times}\usepackage{mathptm}

\begin{document}

\title{ On the {\em ab initio} calculation of CVV Auger spectra in
closed-shell systems }

\author{G. Fratesi,$^{1,2}$ M. I. Trioni,$^{2}$ G. P. Brivio,$^{1,2}$ S. Ugenti,$^{3,4}$ E. Perfetto,$^{1,5}$ and M. Cini$^{3,4}$}

\affiliation{
$^1$Dipartimento di Scienza dei Materiali, Universit\`a di Milano-Bicocca, Via Cozzi 53, 20125 Milano, Italy
\\
$^2$ETSF and CNISM, UdR Milano-Bicocca, Via Cozzi 53, 20125 Milano, Italy
\\
$^3$Dipartimento di Fisica, Universit\`a di Roma Tor Vergata, Via della Ricerca Scientifica 1, I-00133 Roma, Italy
\\
$^4$Laboratori Nazionali di Frascati, Istituto Nazionale di Fisica Nucleare, Via E. Fermi 40, 00044 Frascati, Italy
\\
$^5$Unit\`a CNISM, Universit\`a di Roma Tor Vergata, Via della Ricerca
Scientifica 1, I-00133 Roma, Italy
}

\date{\today} % It is always \today, today,

\begin{abstract}
We propose an {\em ab initio} method to evaluate the
core-valence-valence (CVV) Auger spectrum of systems with filled
valence bands.
The method is based on the Cini-Sawatzky theory, and aims at
estimating the parameters by first-principles calculations in the
framework of density-functional theory (DFT).
Photoemission energies and the interaction energy for the two holes in
the final state are evaluated by performing DFT simulations for the
system with varied population of electronic levels.
Transition matrix elements are taken from atomic results.
The approach takes into account the non-sphericity of the density of
states of the emitting atom, spin-orbit interaction in core and valence,
and non quadratic terms in the total energy expansion with respect to
fractional occupation numbers.
It is tested on two benchmark systems, Zn and Cu metals, leading in
both cases to $L_{23}M_{45}M_{45}$ Auger peaks within $2$~eV from the
experimental ones.
Detailed analysis is presented on the relative weight of the various
contributions considered in our method, providing the basis for
future development.
Especially problematic is the evaluation of the hole-hole interaction
for systems with broad valence bands: our method underestimates its
value in Cu, while we obtain excellent results for this quantity in Zn.
\end{abstract}



\maketitle              %%%%%%%%%%%%%%%%%            TITLE     %%%%%%%%%%%

\section{Introduction}

The Auger spectroscopy involves the creation of two localized holes at
or close to the same atom, hence giving access to local electronic
properties.
Direct information on the local density of valence states is brought by
core-core-valence transitions; in the case of core-valence-valence (CVV)
ones, which will be investigated here, one can in addition access the
screened Coulomb repulsion amongst the two valence holes in the final
state which is relevant to a wide class of phenomena, and study its
effects.

From the theoretical point of view, a large amount of work has been
devoted to the calculation of the Auger spectra of solids during the
last three decades.\cite{Verdozzi:2001:AugerRev} A general formulation
of the dynamical Auger decay, where the creation of the initial core
hole and the Auger decay are considered as coherent processes, was
given by Gunnarsson and
Sch\"{o}nhammer\cite{Gunnarsson:1980:dynTheoryAuger} but is of hard
practical implementation.
In solids with (almost) closed valence bands, where no dynamical
core-hole screening can occur before the Auger decay, one can employ a
simpler two-step approximation and consider the above events as
independent.
Under this assumption, Cini\cite{Cini:1977:CSM} and
Sawatzky\cite{Sawatzky:1977:CSM} (CS) proposed a simple model
providing the Green's function describing the two valence holes left
after the Auger decay.
Good agreement with experiments was achieved using fitting parameters
for the screened Coulomb interaction, giving a quantitative
understanding of the Auger spectra of transition metals located at the
beginning and the end of the row, such as Ti,\cite{Cini:1995:AugerTi}
Ag,\cite{Cole:1994:AgAugerOffSite} and
Au.\cite{Verdozzi:1995:AuAugerOffSite} These results confirmed the
usefulness of including explicitly on-site Hubbard terms to one-body
Hamiltonian, and prompted an extension to nearest-neighbour
interactions.\cite{Verdozzi:1995:OffSiteRev}

These studies determined the relevant physical parameters by
reproducing experimental findings within a semi-empirical approach.
Of particular interest is the parameter governing the interaction
amongst the two holes in the final state, which has an analogue in the
popular LDA+U description for correlated
systems.\cite{Anisimov:1991:LDA+U, Anisimov:1993:LDA+U} Even if
methods for its {\em ab initio} evaluation have been proposed, also
this quantity is often determined by phenomenological arguments.

The possibility of evaluating CVV spectra from first-principles, rather
than from a model with parameters fitted to experiments, would then be
very desirable as it would allow predicting different situations (e.g.,
investigate the effect of a given chemical environment on the Auger
current) and a deeper interpretation of experimental findings.

This paper addresses such possibility, by proposing a
method to compute the parameters entering the CS model by {\em ab
initio} simulations. In this step towards a first-principle
description of CVV Auger spectra in systems where the interaction of
the final-state holes cannot be neglected, we aim at highlighting the
most important contributions to the spectrum, which one should
focus to in forthcoming improvements.
The method is based on Density Functional Theory (DFT) simulations in
the Kohn-Sham (KS) framework, with constrained occupations. We make
use of comparison with reference atomic calculations to extrapolate
the electronic properties of the sample when they are more difficult
to evaluate directly. Results are presented for the
$L_{23}M_{45}M_{45}$ Auger line of Cu and Zn metals, which have been
chosen as benchmark systems with closed $3d$ bands: the former being
more challenging for the proposed procedure, and the latter bearing
more resemblance with the atomic case.

The paper is organized as follows. In section~\ref{sec:methods} we
describe our method to evaluate the Auger spectra by first-principles
calculations.  Section~\ref{sec:results} presents our theoretical
results for Cu and Zn metals, comparing them with experimental results
in the literature.  In section~\ref{sec:discussion} we analyze the
weight of various contributions and discuss improvements.  Finally,
section~\ref{sec:conclusions} is devoted to conclusions.

\section{\label{sec:methods}Theoretical methods}

\subsection{Model Hamiltonian and the Cini-Savatzky solution}

We describe the electron system in the hole representation, by an
Hubbard-like~\cite{Hubbard:1963:HubbardH} model Hamiltonian
\begin{equation}
\label{eq:HHubbard}
H=\epsilon_c c_c^\dagger c_c
 +\sum_v\epsilon_v c_v^\dagger c_v
 +\frac{1}{2}\sum_{\varphi_1 \varphi_2 \varphi_3 \varphi_4} U_{\varphi_1 \varphi_2 \varphi_3 \varphi_4}
c^\dagger_{\varphi_1}c^\dagger_{\varphi_2} c_{\varphi_4}c_{\varphi_3},
\end{equation}
where $c$ and $v$ label the core state involved in the transition and
the valence states of the system, respectively, including the spin
quantum number.  In bulk materials, $v$ is a continuous index.  The
last term is the hole-hole interaction Hamiltonian, parametrized by
the screened repulsion $U$, and is for simplicity restricted here to a
finite set of wavefunctions, $\varphi$, centered at the emitting atom
(hence neglecting interatomic interactions).
In closed shell systems, $\epsilon_c$ and $\epsilon_v$ yield the core
and valence photoemission energies, since the two-body term has no
contribution on the one-hole final state or on the zero-hole initial
one.

A two-step model is adopted to represent the Auger process, assuming
that the initial ionization and the following Auger decay of the core
hole can be treated as two independent events. In other terms, we assume
that the Auger transition we are interested in follows a fully relaxed
ionization of a core shell.
If the ground state energy of the neutral $N$-electron system is chosen
as a reference, the energy of the initial state is simply given by
$\epsilon_c$.
The total spectrum for electrons emitted with kinetic energy $\omega$ is
proportional to
\begin{equation}
\label{eq:Spectrum}
S(\omega)=\sum_{XY} A^{*}_X D_{XY}(\epsilon_c-\omega) A_Y,
\end{equation}
where $X$ and $Y$ are the final-state quantum numbers, $A_X$ is the
Auger matrix element corresponding to the final state $X$, and
$D_{XY}$ represents the two-hole density of states.
Notice that Eq.~(\ref{eq:Spectrum}) coincides with the Fermi golden rule
if the states $X$, $Y$ are eigenstates of the Hamiltonian, so that
$D_{XY}$ is diagonal.
The presence of the transition matrix elements effectively reduces the
set of states contributing to Eq.~(\ref{eq:Spectrum}) to those with a
significant weight close to the emitting atom.
This motivates the approximation to restrict $X$ and $Y$ to two-hole
states based on wavefunctions centered at the emitting atom, such as
the set $\{\varphi\}$ previously introduced.
Therefore, the CVV spectrum is a measure of the two-hole local density
of states (2hLDOS), with modifications due to the matrix elements.
The 2hLDOS could in principle be determined as the imaginary part of
the two-hole Green's function, $G_{XY}$, solution of
Eq.~(\ref{eq:HHubbard}). However, because of the presence of the
hole-hole interaction term, evaluating $G_{XY}$ is in general a
formidable task.

For systems with filled valence bands, the two holes are created in a
no-hole vacuum and one is left with a two-body problem. A solution in
this special case has been proposed by Cini\cite{Cini:1977:CSM} and
Sawatzky,\cite{Sawatzky:1977:CSM} and is briefly reviewed here (see
Ref.~\onlinecite{Verdozzi:2001:AugerRev} for an extended review).
The two-holes interacting Green's function, $G$, is found as the
solution to a Dyson equation with kernel $U$, which reads:
\begin{equation}
\label{eq:G}
G(\omega)=G^{(0)}(\omega) \big(1-UG^{(0)}(\omega)\big)^{-1}.
\end{equation}
Here, $G^{(0)}$ is the non-interacting Green's function which can be
computed from the non-interacting 2hLDOS, $D^{(0)}$, via Hilbert
transform. Such 2hLDOS results from the self-convolution of the
one-hole local density of states (1hLDOS),
$D^{(0)}{\equiv}d{*}d$.

The quantum numbers $LSJM_J$ (intermediate coupling scheme) are the
most convenient choice to label the two-hole states, allowing for the
straightforward inclusion of the spin-orbit interaction in the final
state by adding to the Hamiltonian the usual diagonal term,
proportional to $[J(J+1)-L(L+1)-S(S+1)]$.
Finally, the CVV lineshape is:
\begin{equation}
\label{eq:PCiniLSJMJ}
S(\omega)=-\frac{1}{\pi}
\sum_{LSJM_J \atop L'S'J'M_{J'}'} A^{*}_{LSJ} A_{L'S'J'} \text{Im}
\left[
\frac{
         G^{(0)} ( \epsilon_c - \omega )
}{ %----------------------------------------------------------
    1  -   U G^{(0)}( \epsilon_c - \omega )
}
\right]_{LSJM_J \atop L'S'J'M_{J'}'}.
\end{equation}
For comparison with experimental results, this is to be convoluted
with a Voigt profile to account for core-hole lifetime and
experimental resolution.

It is customary to isolate two limiting regimes:
(i) When $U$ is small with respect to the valence band width $W$
(broad, band-like spectra) the 2hLDOS is well represented by
$D^{(0)}(\omega)$.  However, in such a case it might be even
qualitatively important to account for a dependence of the matrix
elements on the Auger energy $\omega$.  As a consequence, accurate
calculations of the lineshape require the simultaneous evaluation of
the matrix elements and the DOS.
(ii) For $U$ larger than $W$, narrow atomic-like peaks dominate the
spectrum, each peak from an $LSJ$ component.  Hence, to the first
approximation the spectrum is described by a sum of
$\delta$-functions, weighted by matrix elements whose dependence on
the Auger energy may be neglected.  If we take the matrix $U$ diagonal
in the $LSJ$ representation, and indicate by $E^{(0)}_{LSJ}$ the
weighted average of $D^{(0)}_{LSJ}(\omega)$, one obtains:
\begin{equation}
\label{eq:PDeltaLSJ}
S(\omega)\approx\sum_{LSJ}(2J+1)|A_{LSJ}|^2
\delta\big( (\epsilon_c - E^{(0)}_{LSJ} - U_{LSJ}) - \omega \big).
\end{equation}
Atomic matrix elements can be taken as a first approximation, often
satisfactory, and can be evaluated as shown in
Ref.~\onlinecite{Antonides:1977:LMM_Cu_Zn_Ga_Ge_I}.
An approach which could bridge between these two limiting regimes,
considering both finite values of $U$ and the energy dependence of the
matrix elements, is still missing to our knowledge.

In the present work we adopt Eq.~(\ref{eq:PCiniLSJMJ}) in order to
simulate the spectrum. Accordingly, one has to determine the quantities
$A$, $U$, $D^{(0)}(\omega)$, and $\epsilon_c$.
In this paper we make use of a $U$ matrix which does not include the
spin-orbit interaction, and is diagonal on the $LS$ basis.
We take atomic results in the literature for the matrix elements $A$,
which are assumed independent of $J$
too.\cite{Antonides:1977:LMM_Cu_Zn_Ga_Ge_I}
The other quantities are computed by DFT simulations, as detailed in the
following Section.

\subsection{{\em Ab initio} determination of the relevant parameters}

To evaluate {\em ab initio} the photoemission energies we use a method
closely related to Slater's transition-state theory, while the
parameter $U$ is computed following a general procedure first proposed
in Ref.~\onlinecite{Gunnarsson:1989:DFTcalcAnderson} and then adopted
by several authors.

One extrapolates total energies for the system with $N$, $N-1$ and $N-2$
electrons by DFT calculations with constrained occupations for $N-q$
electrons, with $q$ small (typically, up to $0.05$), so that ionized
atoms in otherwise periodic systems can be treated in rather small
supercells. We make the approximation that the total energy of the
system with $q_i$ electrons removed from the level $i$ is given by a
power expansion in $q_i$ up to third order:
\begin{equation}
\label{eq:EtotQ}
E(N-q_i) = E(N) + A_i q_i + B_i q_i^2 + C_i q_i^3.
\end{equation}
In the following we shall assume that this can be extended to finite
values of $q_i$.
The introduction of the cubic term $C_i q_i^3$ allows for a
$q$-dependence of the screening properties of the system.
The coefficients $A_i$, $B_i$, and $C_i$, where $i$ labels core and
valence states involved in the transition, are in this framework all
what is needed to compute the Auger electron energy.  They can be
evaluated in two equivalent ways, whichever is most convenient: by
taking the first, second, and third derivatives of the total energy
$E(N-q_i)$ for $q_i\rightarrow0$; by using Janak's
theorem\cite{Janak:1978:theorem} and computing the KS eigenvalue of
level $i$ and its first and second derivatives:
\begin{equation}
\label{eq:JanakQ}
-\epsilon^\text{KS}_i(N-q_i) = A_i + 2 B_i q_i + 3 C_i q_i^2.
\end{equation}
In particular, $A_i$ is given by (minus) the KS eigenvalue in the
neutral system.

The binding energy of a photoemitted electron, $\epsilon_i
{\equiv}E^\text{XPS}_i=E(N-1_i)-E(N)$, to be used in
Eq.~(\ref{eq:HHubbard}), is given by Eq.~(\ref{eq:EtotQ}) as:
\begin{equation}
\label{eq:XPSABC}
 \epsilon_i = A_i + B_i + C_i.
\end{equation}
This is very close to the well-known Slater's transition-state
approach, in which the XPS energy equals the (minus) eigenvalue at
half filling.  The latter amounts to $A_i+B_i+\frac{3}{4}C_i$ when
approximating the total energy by a cubic expansion as in
Eq.~(\ref{eq:EtotQ}).  In other terms, it differs from the result of
Eq.~(\ref{eq:XPSABC}) only by $\frac{1}{4}C_i$, with $C_i\lesssim1$~eV
in the cases considered here (see below).
It is worth noticing that the term $B_i+C_i$ acts like a correction to
the (minus) KS eigenvalue $A_i$, accounting for dynamical relaxation
effects even though all terms are evaluated within KS-DFT.

The evaluation of the $A$, $B$, and $C$ coefficients for localized
states poses no additional difficulty. Instead, care must be taken
when determining those corresponding to the delocalized valence shells
of bulk materials ($A_v$, $B_v$, and $C_v$), for which we propose the
following method.
As for $A_v$, this is a continuous function of the quantum number $v$
and, by taking advantage of Janak's theorem, it is the KS band energy
with reversed sign.
To estimate $B_v$ and $C_v$, we neglect their dependence on $v$ and
assume that a single value can be taken across the valence band,
acting as a rigid shift of the band.  Hence, the 1hLDOS is obtained
from the KS LDOS, $d^\text{KS}(\omega)$, as:
\begin{equation}
\label{eq:1hLDOS}
d(\omega)=d^\text{KS}(-\omega+B_v+C_v).
\end{equation}
For sake of the forthcoming discussion, one can also define a single
value of $A_v$ in the solid by taking the KS valence band average.

We expect the above approximation to be a good one as long as the
valence band is sufficiently narrow and deep (since eventually the
correction should approach zero at the Fermi level).  Still under this
simplification, the direct evaluation of $B_v$ and $C_v$ would ask for
constraining the occupations for fairly delocalized states, which is a
feasible but uneasy task.
As an alternative route, we suggest a simpler approach based on the
working hypothesis that the environment contribution to the screening
of the positive charge $q_i$ in Eq.~(\ref{eq:EtotQ}) does not depend
strongly on the shape of the charge distribution.
Practically, we take the neutral isolated atom as a reference
configuration in Eq.~(\ref{eq:EtotQ}), and evaluate the coefficients
$B_i^a$ and $C_i^a$ for this system. The two quantities
${\Delta}B=B_i-B_i^a$ and ${\Delta}C=C_i-C_i^a$ can be easily computed
for core levels.
Such bulk-atom corrections are reported in Table~\ref{tab:deltaBC} for
Cu and Zn, which demonstrates that they are almost independent of the
core level. This supports our working hypothesis, and enables us to
extrapolate to the valence shell. Accordingly, $B_v$ and $C_v$ are given
by:
\begin{eqnarray}
\label{eq:B=Ba+DeltaB}
B_v=B_v^a+{\Delta}B,\\
\label{eq:C=Ca+DeltaC}
C_v=C_v^a+{\Delta}C.
\end{eqnarray}

We remark that by choosing the neutral atom as the reference system
some degree of arbitrariness is introduced.  In principle, one could
evaluate the atomic coefficients in a configuration which is
closest to the one of the atom in the solid, depending on its chemical
environment.  However, such arbitrariness has limited effect on the
final value of $B_v$ (similar discussion applies for $C_v$), owing to
cancellations between $B_c^a$ and $B_v^a$ in
Eq.~(\ref{eq:B=Ba+DeltaB}), as will be demonstrated in the following.

\begin{table} %[p!]
\begin{tabular}{|c|c|rrrrr|r|}
\hline
\hline
 &
 & \multicolumn{1}{c}{$1s$}
 & \multicolumn{1}{c}{$2s$}
 & \multicolumn{1}{c}{$2p$}
 & \multicolumn{1}{c}{$3s$}
 & \multicolumn{1}{c|}{$3p$}
 & \multicolumn{1}{c|}{average} \\
\hline
Cu
 & $\Delta{}B$
 & $-4.77$
 & $-4.92$
 & $-4.90$
 & $-4.81$
 & $-4.76$
 & $-4.85\pm0.07$ \\
 & $\Delta{}C$
 & $-0.88$
 & $-0.80$
 & $-0.82$
 & $-0.73$
 & $-0.72$
 & $-0.81\pm0.07$ \\
\hline
Zn
 & $\Delta{}B$
 & $-4.21$
 & $-4.26$
 & $-4.27$
 & $-4.19$
 & $-4.17$
 & $-4.23\pm0.04$ \\
 & $\Delta{}C$
 & $-0.49$
 & $-0.33$
 & $-0.26$
 & $-0.32$
 & $-0.35$
 & $-0.35\pm0.09$ \\
\hline
\hline
\end{tabular}
\caption{\label{tab:deltaBC}
Differences amongst the values of $B$ and $C$ in the bulk and the atom,
$\Delta{}B=B-B^a$ and $\Delta{}C=C-C^a$, for core levels of Cu and Zn.
The last column reports the average and standard deviation across the
core levels.
All values in eV.
}
\end{table}

Regarding the interaction energy $U$ for the two holes in a valence
level, defined by $[E(N-2)-E(N)]-2[E(N-1)-E(N)]$, let us consider the
case of spherically symmetric holes (non-spherical contributions,
giving rise to multiplet splitting, will then be added).  Such
spherical interaction, denoted by $U_\text{sph}$, can be determined
via Eq.~(\ref{eq:EtotQ}), resulting in
\begin{equation}
U_\text{sph} = 2B_v + 6C_v.
\end{equation}
This amounts to the second derivative of the DFT energy as a function
of the band occupation,
$U(q)= \partial^2 E(N-q) / \partial q^2 $,
as originally suggested by Gunnarsson and
coworkers,\cite{Gunnarsson:1989:DFTcalcAnderson} here evaluated for
the $N-1$-electron system rather than for the neutral one.
Differently, the interaction energy commonly used in LDA+U
calculations of the ground state is defined as $E(N+1)+E(N-1)-2E(N)$
and hence evaluated by the second derivative in $q=0$, resulting in
$2B_v$ only.
Notice here that the role of the cubic term in Eq.~(\ref{eq:EtotQ}) is
to introduce a dependence of the interaction energy on the particle
number, following the one of the screening properties of the system.
Finally, non-spherical contributions, which give rise to multiplet
splitting, are added to $U_\text{sph}$. It has been
demonstrated\cite{Antonides:1977:LMM_Cu_Zn_Ga_Ge_I} for a number of
materials, including Cu and Zn, that these terms are well reproduced
by a sum of atomic Slater integrals,\cite{Slater:1960:atoms}
$a_2F^2+a_4F^4$, where the coefficients $a_2$ and $a_4$ depend on the
multiplet configuration and
\begin{equation}
F^k=\int_0^\infty r_1^2 dr \int_0^\infty r_2^2 dr_2
    \frac{r^k_<}{r^{k+1}_>}
    \left[\varphi^a(r_1)\varphi^a(r_2)\right]^2.
\end{equation}
Here $\varphi^a(r)$ is the atomic radial wave function relevant to the
process under investigation (e.g., the $3d$ one for a $CM_{45}M_{45}$
Auger transition), and $r_<$ ($r_>$) is the smaller (larger) of $r_1$
and $r_2$.
Notice that the spherical Slater integral $F^0$ is implicit into
$U_\text{sph}$, which has the meaning of a screened Coulomb
integral.\cite{Anisimov:1991:F0effDFT}

Summarizing, one has:
\begin{equation}
\label{eq:U}
U=2B_v^a+6C_v^a+2{\Delta}B+6{\Delta}C+a_2F^2+a_4F^4.
\end{equation}
It is customary to write $U=F-R$, where $F=F^0+a_2F^2+a_4F^4$, and $R$
is the ``relaxation energy''.\cite{Shirley:1973:KLL_relaxation} This can
be further decomposed into an atomic and an extra-atomic contribution,
$R=R_a+R_e$.  From Eq.~(\ref{eq:U}), one identifies
$R_a=F^0-2B_v^a-6C_v^a$ and $R_e=-2{\Delta}B-6{\Delta}C$.  Notice that by
our approach we compute $F^0-R_a$ as a single term, so that it is not
possible to separate the two contributions.

In other formulations,\cite{Pickett:1998:LDA+U, Cococcioni:2005:LDA+U}
the derivative of the energy with respect to the occupation number of
a broad band is computed by shifting the band with respect to the
Fermi level.  This adds a non-interacting contribution to the
curvature of the energy, since the level whose occupation is varied is
itself a function of the band occupancy.  Such non-interacting term
has to be subtracted when computing $U$ by these
approaches.\cite{Cococcioni:2005:LDA+U} Our formulation is
conceptually more similar to scaling the occupation of all valence
atomic levels in a uniform way, and the non-interacting term is
vanishing.

\subsection{Computational details}

The results presented in this paper have been obtained by DFT
calculations with the Perdew-Burke-Ernzerhof\cite{Perdew:1996:PBE}
generalized gradient approximation for the exchange and correlation
functional.
We used an all-electron linearized augmented-plane-wave code to perform
the simulations with constrained core occupations. Periodically repeated
supercells at the experimental lattice constants were adopted to
describe the solids.
One atom was ionized in a unit cell containing four and eight atoms
for Cu and Zn, respectively.  In both cases the ionized atom has no
ionized nearest neighbours.
Cell neutrality is preserved by increasing the number of the valence
electrons, simulating the screening of the core hole by the solid.
The spin-orbit splitting in core states as well as in the final state
with two holes was taken into account by adopting DFT energy shifts
for free atoms,~\cite{NIST::DFTdata} and is here assumed independent
on the fractional charge $q$ (we verified that the latter approximation
affects our final results by no more than $0.2$~eV).
As for the coefficients $A$, $B$, and $C$ in Eq.~(\ref{eq:EtotQ}), we
found values numerically more stable, with respect to convergence
parameters, by performing a second order expansion of the eigenvalues
rather than a third order expansion of the total energy.  Therefore,
we made use of Janak's theorem and Eq.~(\ref{eq:JanakQ}), with
eigenvalues relative to the Fermi level in the solid (hence, resulting
XPS and Auger energies are given with respect to the same reference).
Fulfillment of Janak's theorem and coincidence of results of
Eq.~(\ref{eq:EtotQ}) and (\ref{eq:JanakQ}) were numerically verified
to high accuracy in a few selected cases.  The values of $q$ ranged
from $0$ to $0.05$ at intervals of $0.01$.
Comparison with denser and more extended meshes for the free atom case
showed that results are not dependent on the chosen mesh.  Matrix
elements and Slater integrals $F^2$ and $F^4$ are taken from
Ref.~\onlinecite{Antonides:1977:LMM_Cu_Zn_Ga_Ge_I}, and core hole
lifetimes from Ref.~\onlinecite{Yin:1973:widthL2L3}.

\section{\label{sec:results}Results}

In this Section we report our results for the $L_{23}M_{45}M_{45}$
Auger lineshape of Cu and Zn.  The core and the valence indices, $c$
and $v$ in the previous Section, are specialized to the $2p$ and $3d$
level of such elements, respectively.

\begin{figure} %[p!]
\includegraphics[width=8.5cm]{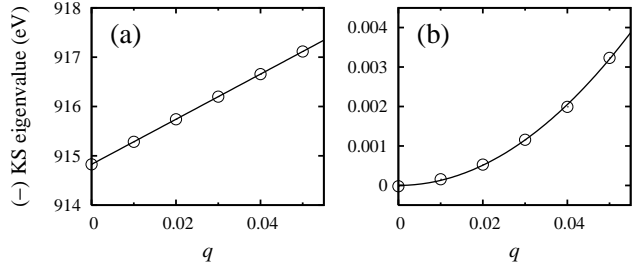}
\caption{\label{fig:sampleFit}Example of fitting the Kohn-Sham
eigenvalue to extract the coefficients $A$, $B$, and $C$.  Results are
shown for the $2p$ level of metal Cu.  Panel (a) plots the KS
eigenvalue (relative to the Fermi energy) with reversed sign (circles)
and the fitted parabola from Eq.~(\ref{eq:JanakQ}) (line), as function
of the number of electrons removed from the $2p$ level, $q$. Panel (b)
reports the same quantities after subtracting the linear term
$A+2Bq$.}
\end{figure}

As an example of our procedure to extract the parameters $A$, $B$, and
$C$ [see Eq.~(\ref{eq:EtotQ})], we report the case for the $2p$ level
of Cu metal in Fig.~\ref{fig:sampleFit} (the following considerations
are also valid in the other cases).
We remove the fractional number of electrons $q$ from the $2p$ level
of a Cu atom, and plot its (minus) KS $2p$ eigenvalue in
Fig.~\ref{fig:sampleFit}a.
Such a curve is fitted by the expression in Eq.~(\ref{eq:JanakQ}).
It is apparent from Fig.~\ref{fig:sampleFit}a that a linear fit
already reproduces the KS eigenvalue in this range of $q$ to high
accuracy.  However, since results are to be extracted up to $q=1$ or
$2$, the quadratic term in the expansion is also of interest.  This is
shown in Fig.~\ref{fig:sampleFit}b, where the linear contribution
($A+2Bq$) has been subtracted. The parabola accurately fits the
numerical results, with residuals of the order of $10-50$~$\mu$eV.

Table~\ref{tab:ABC_CuZn} collects our results for the coefficients
$A$, $B$, and $C$, needed for the determination of the
$L_{23}M_{45}M_{45}$ lineshape of Cu and Zn.
The values of $B$ and $C$ for the $L_2$ and $L_3$ cases are identical,
following the assumption that spin-orbit splitting is independent of
the fractional charge.
The coefficients $B_{M_{45}}$ and $C_{M_{45}}$ in the solid have been
obtained by comparing results for core levels in the bulk and in the
free neutral atom according to
Eqs.~(\ref{eq:B=Ba+DeltaB}-\ref{eq:C=Ca+DeltaC}), with the values of
$\Delta{}B$ and $\Delta{}C$ averaged across the core levels as
reported in Table~\ref{tab:deltaBC}.
Their negative sign indicates that the interaction amongst the two
holes is more effectively screened in the solid.

\begin{table} %[p!]
\begin{tabular}{|c|c|rrr|rrr|rr|}
\hline
\hline
\multicolumn{2}{|c|}{Level} & \multicolumn{1}{c}{$A^a$}          & \multicolumn{1}{c}{$B^a$}
& \multicolumn{1}{c|}{$C^a$}  &
        \multicolumn{1}{c}{$A$}            & \multicolumn{1}{c}{$B$}
& \multicolumn{1}{c|}{$C$}    &
        \multicolumn{1}{c}{$E^\text{XPS}$} & \multicolumn{1}{c|}{Exp.}
\\
\hline
   &$L_2$    & $ 930.17$ & $27.74$ & $1.25$   &   $ 928.40$ & $22.84$ & $0.43$   &   $ 951.67$ & $ 952.0$ \\
Cu &$L_3$    & $ 909.81$ & $27.74$ & $1.25$   &   $ 908.04$ & $22.84$ & $0.43$   &   $ 931.31$ & $ 932.2$ \\
   &$M_{45}$ & $   5.04$ & $ 5.72$ & $0.91$   &   $   2.86$ & $ 0.87$ & $0.10$   &   $   3.84$ & $   3.1$ \\
\hline
   &$L_2$    & $1019.40$ & $30.44$ & $1.08$   &   $1016.98$ & $26.17$ & $0.82$   &   $1043.97$ & $1044.0$ \\
Zn &$L_3$    & $ 995.69$ & $30.44$ & $1.08$   &   $ 993.27$ & $26.17$ & $0.82$   &   $1020.26$ & $1020.9$ \\
   &$M_{45}$ & $  10.14$ & $ 7.06$ & $0.77$   &   $   7.53$ & $ 2.82$ & $0.42$   &   $  10.78$ & $   9.9$ \\
\hline
\hline
\end{tabular}
\caption{\label{tab:ABC_CuZn}
Coefficients for the expansion of the total energy as a function of
the number of electrons, $E(N-q)$, for atomic ($A^a$, $B^a$, $C^a$)
and bulk ($A$, $B$, $C$) Cu and Zn.
As for the $M_{45}$ values: by $A_{M_{45}}$ we indicate (minus) the
weighted average of the $3d$ KS band; $B_{M_{45}}$ and $C_{M_{45}}$
are obtained according to
Eqs.~(\ref{eq:B=Ba+DeltaB}-\ref{eq:C=Ca+DeltaC}).  Theoretical XPS
energies are given by Eq.(\ref{eq:XPSABC}); experimental data are
taken from Ref.~\onlinecite{Antonides:1977:LMM_Cu_Zn_Ga_Ge_I}.
Values in eV.
}
\end{table}

Let us now consider the dependence of our results on the particular
choice of the reference atomic configuration.
For comparsion, the Cu$^+$ and Zn$^+$ ions (with one electron removed
from the $4s$ shell) have been used as a starting point for the
evaluation of the atomic coefficients instead of the neutral one.
We find similar modifications, canceling each other in
Eq.~(\ref{eq:B=Ba+DeltaB}), for core and valence $B_i^a$ atomic
coefficients (larger by about $1.5$~eV in Cu and $1.3$~eV in Zn).  The
same is found for the $C_i^a$ coefficients (lower by $0.2$~eV in Cu
and $0.1$~eV in Zn).
As a consequence, the values for $B_{M_{45}}$ differ by less than
$0.2$~eV, and those for $C_{M_{45}}$ are identical within $0.01$~eV,
with data reported in Table~\ref{tab:ABC_CuZn}.
Hence, as anticipated in the previous section, the choice of the
reference atomic configuration does not affect significantly the
evaluated XPS and Auger energies.

Recall now that $A+B+C$ is our estimate for the XPS excitation
energies [see Eq.~(\ref{eq:XPSABC})], which are reported in
Table~\ref{tab:ABC_CuZn}, and compared with experimental
values.\cite{Antonides:1977:LMM_Cu_Zn_Ga_Ge_I}
Notice that bare KS excitations energies can be $30$~eV smaller than
the experimental value, but the addition of $B$ and, to a smaller
extent, of $C$, properly accounts for the missing relaxation energy,
the left discrepancy being smaller that $1$~eV.

\begin{figure} %[p!]
\includegraphics[width=8.5cm]{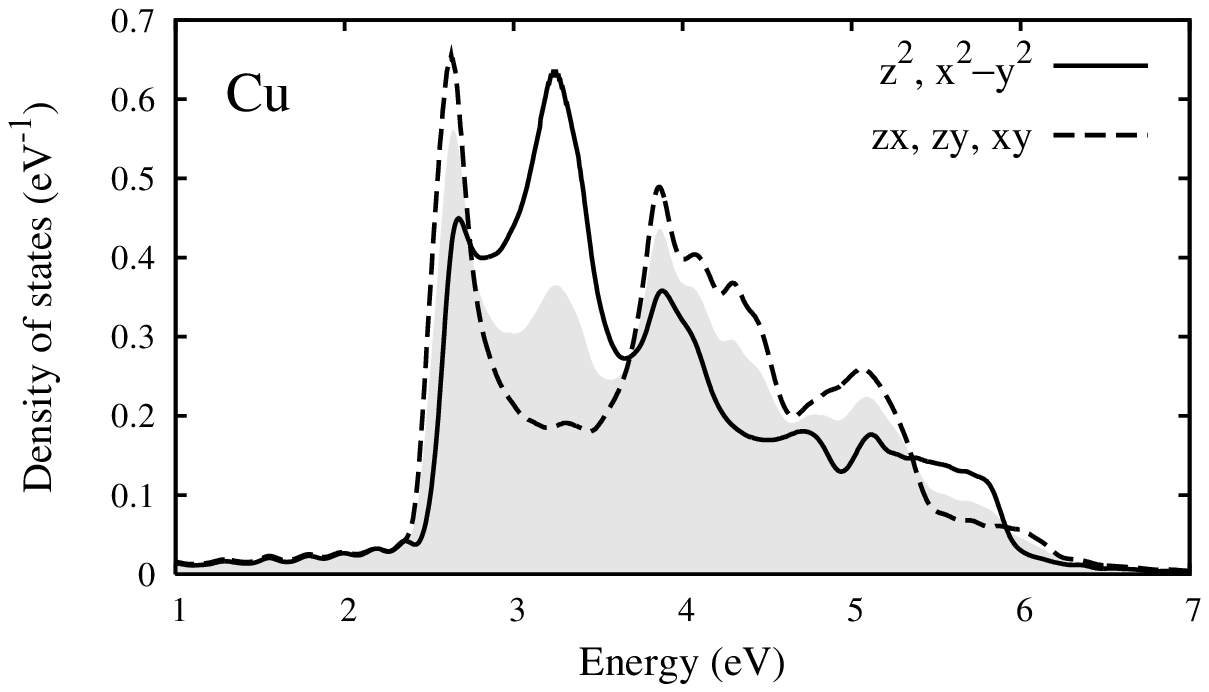}
\includegraphics[width=8.5cm]{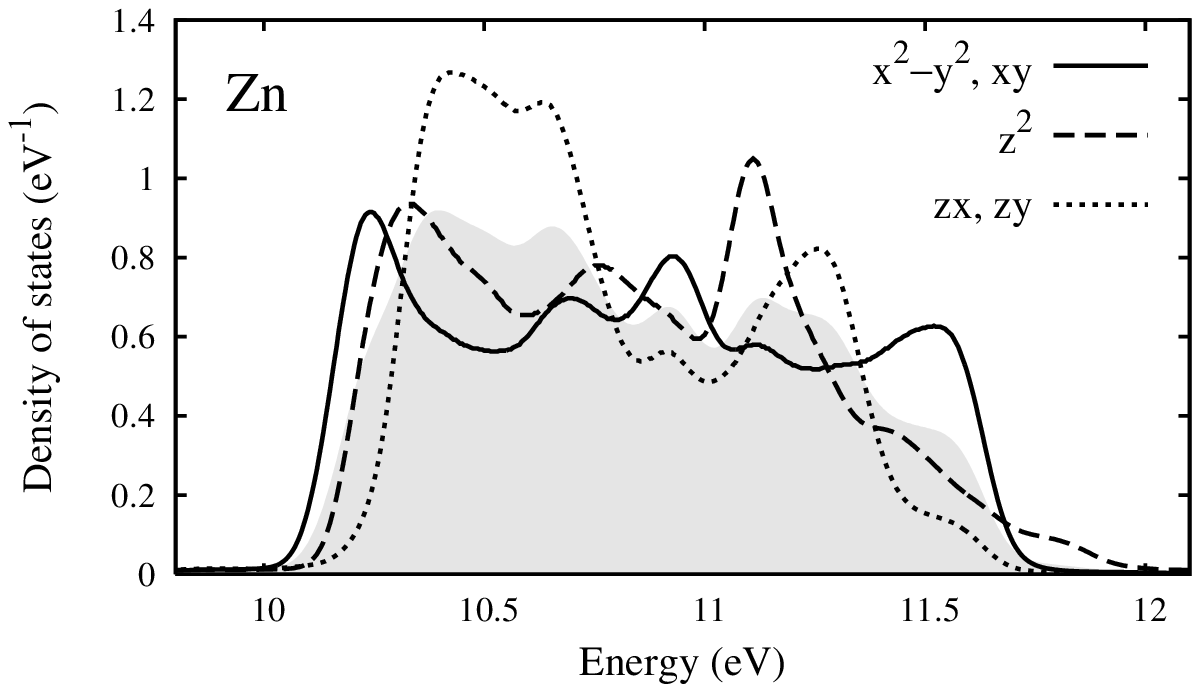}
\caption{\label{fig:1hLDOS_CuZn}One-hole local density of states,
$d(\omega)$, for Cu (top) and Zn (bottom), relative to the Fermi
energy and normalized to unity.  The shaded area is the total $d$ DOS,
$\bar{d}$.}
\end{figure}

We report next our results for the $3d$ component of the 1hLDOS,
$d(\omega)$, for Cu and Zn in Fig.~\ref{fig:1hLDOS_CuZn}.  We remind
that such quantity is obtained by converting the KS density of states
into the hole picture, and by translating the result by $B_v+C_v$ to
account for relaxation effects [see Eq.~(\ref{eq:1hLDOS})].
The total $d$ 1hLDOS, $\bar{d}(\omega)$, is shown as a shaded area
together with the components on the different irreducible
representations over which the $d$ matrix is diagonal. For both
metals, the various components differ among themselves in the detailed
energy dependence, but their extrema are very similar.

As a final ingredient, Table~\ref{tab:U} lists the values of $U$ for
the five $LS$ components of the multiplet, computed by
Eq.~(\ref{eq:U}).  Notice that the inclusion of a $q$ dependence in
$U$ (via a cubic term in the expansion of the total energy with
respect to a fractional charge) proves to be quite important.
Indeed, in evaluating $U$ the $C$ coefficient is counted six times,
hence bringing a larger contribution than in the XPS energies
previously discussed.
Such inclusion gives an estimate of $U=U(q=1)$ which is $0.60$~eV and
$2.52$~eV larger for Cu and Zn, respectively, than the corresponding
values obtained as $U(q=0)$.

\begin{table} %[p!]
\begin{tabular}{|c|r|rrrrr|}
\hline\hline
 & \multicolumn{1}{c|}{$U_\text{sph}$}
 & \multicolumn{1}{c}{$^1S$}
 & \multicolumn{1}{c}{$^1G$}
 & \multicolumn{1}{c}{$^3P$}
 & \multicolumn{1}{c}{$^1D$}
 & \multicolumn{1}{c|}{$^3F$} \\
\hline
Cu & 2.38
&7.76
&3.34
&2.67
&2.25
&0.33
\\
\hline
Zn & 8.16
&14.31
&9.26
&8.49
&8.01
&5.82
\\
\hline
\hline
\end{tabular}
\caption{\label{tab:U} Values of $U$ resulting from the application of
Eq.~(\ref{eq:U}), in eV.  Slater's integrals from
Ref.~\onlinecite{Antonides:1977:LMM_Cu_Zn_Ga_Ge_I}.}
\end{table}

\begin{figure} %[p!]
\includegraphics[width=8.5cm]{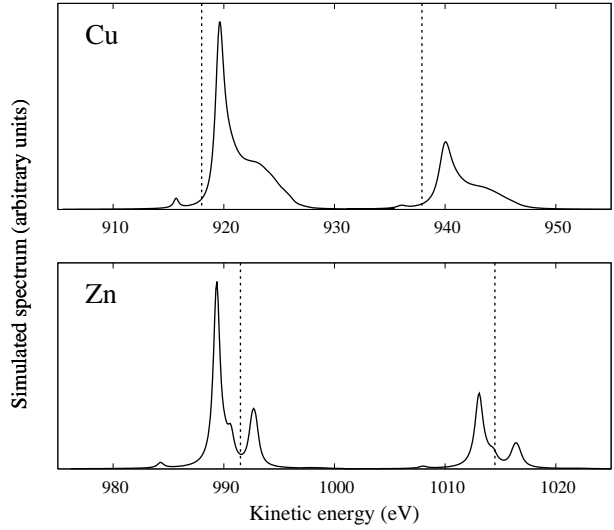}
\caption{\label{fig:LVV}Simulated $L_{23}M_{45}M_{45}$ spectrum for Cu
(top) and Zn (bottom) metals.  The vertical lines mark the position of
the principal ($^1G$) peaks from
experiments.\cite{Antonides:1977:LMM_Cu_Zn_Ga_Ge_I}}
\end{figure}

We then compute the $L_{23}M_{45}M_{45}$ Auger spectrum following
Eq.~(\ref{eq:PCiniLSJMJ}).
The outcome has been convoluted with a core hole lifetime of $0.49$
and $0.27$~eV ($0.42$ and $0.33$~eV)\cite{Yin:1973:widthL2L3} for the
$L_2$ and $L_3$ lines of Cu (Zn), respectively, and results in a
multiplet of generally narrow atomic-like peaks, shown in
Fig.~\ref{fig:LVV}.

To analyze these results, let us focus on the principal peak ($^1G$) in
the spectrum, which can be associated with the absolute position of the
multiplet. (The internal structure of the multiplet in our description
only depends on the values of $F^2$ and $F^4$ which, as previously
specified, were taken from the literature.)
The experimental energy of the (most intense) $^1G$
transition\cite{Antonides:1977:LMM_Cu_Zn_Ga_Ge_I} is marked by a
vertical line in Fig.~\ref{fig:LVV}.  The agreement of our results is
rather good considering the absence of adjustable parameters in the
theory: focusing on the $L_3VV$ line, the $^1G$ peak position
($918.0$~eV from experiments) is overestimated by $1.6$~eV, while the
one for Zn ($991.5$~eV) is underestimated by $1.9$~eV.

\section{\label{sec:discussion}Discussion}

It is interesting to compare these results with those obtained by an
expression commonly adopted for Auger energies, i.e., $\omega_{LS}
\approx \epsilon_c - 2\epsilon_v - U_{LS}$.  This is an excellent
approximation when $U$ is larger than $W$: for example in Zn, where
$W\approx1.5$~eV and $U_{^1G}=9.26$~eV, its application to the
computed parameters yields a value which is only $0.10$~eV larger than
the $^1G$ peak position derived from Eq.~(\ref{eq:PCiniLSJMJ}).
However, when $U$ is of order of $W$, significant deviations can be
observed: e.g., for Cu ($W\approx3.5$~eV and $U_{^1G}=3.34$~eV) the
$^1G$ position is overestimated by $0.66$~eV.  For smaller values of
$U$, the quasi-atomic peak is lost for a broad band-like structure.
This is the case for the $^3F$ component of Cu (the rightmost shoulder
in the spectrum) for which we obtain $U_{^3F}=0.33$~eV.  However, this
is an artifact of our underestimate of $U_\text{sph}$ in Cu:
experimentally, the $^3F$ peak is resolved as well.

Besides these observations, the expression $\omega_{LS} \approx
\epsilon_c - 2\epsilon_v - U_{LS}$ is accurate enough for the $^1G$
peak to discuss the discrepancy of our results with respect to the
experimental ones.
Let us focus on the $L_3VV$ part of the spectrum, and decompose the
Auger kinetic energy $\omega$ into its contributions (see
Table~\ref{tab:decomposeEkin}).
Despite the fact that the overall agreement is similar in magnitude for
Cu and Zn, it is important to remark that this finding has different
origins.
In both metals, we underestimate slightly the core photoemission
energy and overestimate the valence photoemission energy by a similar
amount.  Both effects contribute underestimating the kinetic energy.
In Zn, where our value of $U$ is excellent, the error in $\omega$ stems
from the errors in the photoemission energies.
In Cu, instead, $U$ is seriously underestimated.  This overcompensates
the error in the photoemission energies, resulting in a fortuitous
overall similar accuracy.

\begin{table} %[p!]
\begin{tabular}{|c|c|rrr|r|}
\hline
\hline

 &
 & \multicolumn{1}{c}{$\epsilon_c$}
 & \multicolumn{1}{c}{$-2\epsilon_v$}
 & \multicolumn{1}{c|}{$-U_{^1G}$}
 & \multicolumn{1}{c|}{$\omega_{^1G}$} \\
\hline %           Ec        -2EV        -U      omega
    & Theory &  $931.31$ &  $-7.67$ & $-3.34$ & $920.29$ \\
Cu  & Exp.   &  $932.2$  &  $-6.2$  & $-8.0$  & $918.0$  \\
    & Diff.  &   $-0.9$  &  $-1.5$  &  $4.7$  &   $2.3$  \\
\hline %           Ec        -2EV        -U      omega
    & Theory & $1020.26$ & $-21.55$ & $-9.26$ & $989.45$ \\
Zn  & Exp.   & $1020.9$  & $-19.8$  & $-9.5$  & $991.5$  \\
    & Diff.  &   $-0.6$  &  $-1.8$  &  $0.2$  &  $-2.1$  \\
\hline
\hline
\end{tabular}
\caption{\label{tab:decomposeEkin}
Decomposition of the $^1G$ $L_3M_{45}M_{45}$ Auger kinetic energy into
its contributions, according to the simple approximation
$\omega=\epsilon_c-2\epsilon_v-U$.  Theoretical values of
$\epsilon_c$, $\epsilon_v$, and $U$ from tables \ref{tab:ABC_CuZn} and
\ref{tab:U}; experimental data from
Ref.~\onlinecite{Antonides:1977:LMM_Cu_Zn_Ga_Ge_I}.
Values in eV.
}
\end{table}

We now examine the relative weight of two different ingredients of our
method.  First, the role of the spin-orbit interaction in the final
state, which will be analyzed by comparing with results where such
term is neglected; second, the resolution of the 2hLDOS in its angular
components.  To this respect, we notice that the matrix expression for
the spectrum given in Eq.~(\ref{eq:PCiniLSJMJ}) can be significantly
simplified under the assumption that the 2hLDOS is spherically
symmetric and the spin-orbit contribution can be neglected. In this
case, we can just take the spherically averaged, i.e., the total $d$
1hLDOS, and compute its self-convolution,
$\bar{D}^{(0)}\equiv\bar{d}*\bar{d}$.
An averaged Green's function, $\bar{G}^{(0)}$, is then defined as the
Hilbert transform of $\bar{D}^{(0)}$. By replacing
$G^{(0)}_{LSJM_J,L'S'J'M_{J'}'}$ in Eq.~(\ref{eq:PCiniLSJMJ}) with the
diagonal matrix $\delta_{LSJM_J,L'S'J'M_{J'}'}\bar{G}^{(0)}$, we
obtain the simple scalar equation
\begin{equation}
\label{eq:PCini0}
S(\omega)=-\frac{1}{\pi}\sum_{LSJ} (2J+1)|A_{LS}|^2 \text{Im}
\left[
\frac{
         \bar{G}^{(0)} ( \epsilon_c - \omega )
}{ %-----------------------------------------------------------
    1  -   U_{LS} \bar{G}^{(0)}( \epsilon_c - \omega )
}
\right],
\end{equation}
where the dependence on $LS$ quantum numbers is only via the matrix
elements and the interaction matrix $U$, and each $LS$ component of
the spectrum is decoupled from the others.

\begin{figure} %[p!]
\includegraphics[width=8.5cm]{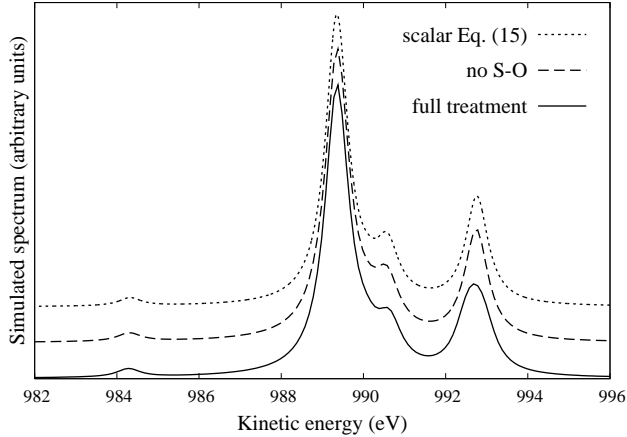}
\caption{\label{fig:SOvsALL}Simulated spectrum for the
$L_3M_{45}M_{45}$ line of Zn.  Solid line: full treatment of
Eq.~(\ref{eq:PCiniLSJMJ}), as presented in this paper. Dashed line:
neglecting the spin-orbit interaction in the two-hole final
state. Dotted line: adopting the spherically averaged 2hLDOS and the
scalar formulation, Eq.~(\ref{eq:PCini0}).  The origin of the vertical
axis is shifted for improved clarity.}
\end{figure}

\begin{figure} %[p!]
\includegraphics[width=8.5cm]{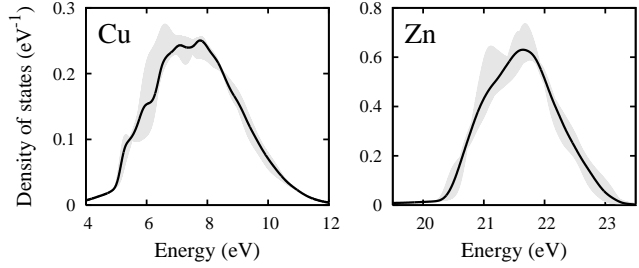}
\caption{\label{fig:2hLDOS_CuZn}Non-interacting two-holes density of
states, $D^{(0)}$, for Cu (left) and Zn (right).  The solid curve
represents $\bar{D}^{(0)}\equiv\bar{d}*\bar{d}$, the self-convolution
of the total $d$ 1hLDOS; the shaded area indicates the largest
deviations from this result found amongst the angular resolved
2hLDOSs.}
\end{figure}

The Auger spectra simulated neglecting the spin-orbit interaction and
calculated with the simplified expression of Eq.~(\ref{eq:PCini0}) are
plotted in Fig.~\ref{fig:SOvsALL} as dashed and dotted line,
respectively, to be compared with the result of the full calculation
[Eq.~(\ref{eq:PCiniLSJMJ})], solid line.
For simplicity, we limit the discussion to the $L_3VV$ line of Zn. The
resemblance of the three results is remarkable. Indeed, in Cu and Zn
the spin-orbit splitting for $3d$ levels is relatively small, $0.27$
and $0.36$~eV, respectively.\cite{NIST::DFTdata} Furthermore, despite
the differences which characterize the angular components of the
1hLDOS (see Fig.~\ref{fig:1hLDOS_CuZn}), the convoluted 2hLDOS are
only mildly different from $\bar{D}^{(0)}$, as reported in
Fig.~\ref{fig:2hLDOS_CuZn}.
Now, in systems with a large $U/W$ ratio, fine details of the 2hLDOS
are not relevant for the position of quasi-atomic peaks, which only
depend on the weighted averages $E^{(0)}_{LS}$ as in
Eq.~(\ref{eq:PDeltaLSJ}).  In our case, the values of $E^{(0)}_{LS}$
lie within $0.1$~eV from those corresponding to the averaged
2hLDOS. Hence the practically equivalent results obtained by
Eq.~(\ref{eq:PCiniLSJMJ}) and Eq.~(\ref{eq:PCini0}).

This analysis shows that, for a wide class of systems with strong
hole-hole interaction, weak spin-orbit interaction, and a spherical
symmetry to some extent, the simple formulation presented in
Eq.~(\ref{eq:PCini0}) is practically as accurate as the expression in
Eq.~(\ref{eq:PCiniLSJMJ}). One should instead adopt the full treatment
for, e.g., heavier elements, or systems with low dimensionality.
This remark is independent of the methodology to determine the
parameters entering the model, either fully {\em ab initio} as in the
present approach, or by phenomenological arguments.

Our method provides an agreement with experimental photoemission
energies of the order of $1$~eV and of $2$~eV for Auger energies.
We consider this to be a rather good result, as a starting point,
considering the absence of adjustable parameters in the model which is
the new feature of our approach for CVV transitions in correlated
systems.
We trust that our simple method could already yield qualitative
information on the variations of the spectrum to be expected following
to modifications in the sample, e.g., when the emitting atom is
located in different environments.
Of course, a much better agreement would be obtained by inserting
phenomenological parameters, but at the cost of loosing predictive
power.

The internal structure of the lineshape, i.e., the multiplet
splitting, is given very precisely.  However, the first-principles
treatment of the latter is not a new aspect of our approach, which is
indeed based in this respect on atomic results available in the
literature since decades.
Let us instead focus again on the estimated position of the multiplet,
which crucially depends on the parameters evaluated by our {\em ab
initio} method.
Even though the agreement with the experiment is about as good as for
Zn as for Cu, actually the results for Zn are much better.
In Zn, the $3d$ band is deep and narrow, and electronic states bear
mostly atomic character. Their hybridization with the states closer to
the Fermi level, which mainly contributes to screening, is small,
somehow in an analogous way as for core states. Consequently, the
approximations to neglect the energy dependence of $B_v$ and $C_v$,
and to transfer the values of $\Delta{}B$ and $\Delta{}C$ from the
core states to the valence ones, produce very good results.
In Cu, instead, the $3d$ band is higher and broader, and hybridizes
significantly with the $s$-like wavefunctions.  Our approximations
turn out to be less adequate: the resulting $U$ is about half the one
derived from experiments.

Part of this discrepancy might also have a deeper physical origin,
since the Hamiltonian adopted, Eq.~(\ref{eq:HHubbard}), does not allow
for the interaction amongst two holes located at different atomic
sites.  The CS model has been extended to consider the role of
interatomic (``off-site'') correlation effects which mainly produce an
energy shift of the Auger line to lower
energies.\cite{Verdozzi:1995:OffSiteRev} Studies based on
phenomenological parameters suggest that such an energy shift could be
of about $2.5$~eV in Cu\cite{Ugenti:2008:spinAPECS} (smaller values
are expected in Zn where holes are more localized and screening is
more effective).  For sake of simplicity, the off-site term has not
been considered here and is left for future investigations.  The
parameters entering this term could be determined by {\em ab initio}
methods in analogy to the procedure shown here for evaluating $U$. It
is however important to notice that adding the off-site term would not
fix all the discrepancies observed in Cu, where also the lineshape, in
addition to the peak position, is not satisfactory owing to small
values for the on-site interaction $U$ (e.g., the non-resolved $^3F$
peak).

Enhancing the accuracy of the values of $U$ seems therefore the most
important improvement for the method presented here, especially for
systems with broad valence bands.  As a possibility, it would be
interesting to use approaches which are capable to compute the total
energy in presence of holes in the valence state. The methodology
presented in Ref.~\onlinecite{Cococcioni:2005:LDA+U}, in which the
valence occupation is changed by means of Lagrange multipliers
associated with the KS eigenvalues, could be particularly effective.
One should pay attention as some arbitrariness is anyway introduced.
Namely, the value of $U$ does depend on the chosen form of the valence
wavefunctions.  Such an arbitrariness is compensated in LDA+U
calculations performed self-consistently.\cite{Cococcioni:2005:LDA+U}
Furthermore, to apply this method to systems with closed band lying
well below the Fermi energy, large shifts of the KS eigenvalues would
be needed to alter the occupation of the band to an appreciable
amount.

Another possible improvement concerns the photoemission energies.
Calculations by the $GW$ approach\cite{Aryasetiawan:1998:GW} of the
1hLDOS could be used to account for relaxation energies, rather than
adopting Eq.~(\ref{eq:XPSABC}).  Results available in the literature
(e.g., for Cu~\cite{Marini:2002:GWCu}) are very promising in that
sense.
It is interesting to notice that the factor $B+C$ plays the role of a
self-energy expectation value, and that the use of a single value of
$B+C$ to shift rigidly the band is formally analogous to the ``scissor
operator'' often introduced to avoid expensive self-energy calculations.
The accuracy of such rigid shifts for valence-band photoemission in Cu
is discussed in Ref.~\onlinecite{Marini:2002:GWCu}.

Systems with larger band width or smaller hole-hole interaction would
require to extend the approach to treat the dependence of the matrix
elements on energy together with the interaction in the final state.
Releasing the assumption that matrix elements equal the atomic ones,
as in the current treatment, or that particles are non-interacting in
formulations accounting for such an energy dependence (like, e.g., the
one in Ref.~\onlinecite{Bonini:2003:MDS}), would allow switching
continously between systems with band-like and atomic-like spectra.
This possibility is currently under investigation.

Finally, let us recall the basic assumption considered here that the
valence shell is closed, which is crucial to the CS model in its
original form.  Efforts have been devoted towards releasing this
assumption, resulting in a formulation by more complicated three-hole
Green's functions,\cite{Marini:1999:3bodyCVVAuger} for which no {\em
ab initio} treatment is nowadays available to our knowledge.

\section{\label{sec:conclusions}Conclusions}

We have presented an {\em ab initio} method for computing CVV Auger
spectra for systems with filled valence bands, based on the
Cini-Sawatzky model.
Only standard DFT calculations are required, resulting in a very
simple method which allows working out the spectrum with no adjustable
parameters. The accuracy on the absolute position of the Auger
features is estimated to a few eV, as we have demonstrated by the
analysis of Cu and Zn metals.
We have shown that in these systems further simplifications like
neglecting spin-orbit interaction for the two valence holes, or the
non sphericity of the emitting atom, give results practically
equivalent to the full treatment.
Attention should be paid to the problematic parameter $U$.  We
obtained such a term with a good accuracy for the more localized,
atomic-like valence bands in Zn, while it results underestimated in
Cu.
Its occupation number dependence, included via a cubic term in the
expansion of the total energy, has been considered, and shown to play
an important role.

This step towards a first-principles description of CVV spectroscopy
in closed-shell correlated systems enables identifying improvements
which future investigations could focus on. In particular, one would
benefit from detailed calculations of the single-particle densities of
states (e.g., by the $GW$ method), from truly varying the valence-band
occupation to obtain the $U$ parameter, and from including off-site
terms in the Hamiltonian. Prospectively, it would be desirable to take
into account the energy dependence of the transition matrix elements.

\section{Acknowledgment}

This work was supported by the MIUR of Italy (Grant
No. 2005021433-003) and the EU Network of Excellence NANOQUANTA (Grant
No. NMP4-CT-2004-500198).  Computational resources were made available
also by CINECA through INFM grants.  EP is financially supported by
Fondazione Cariplo (n. Prot. 0018524).

\bibliographystyle{apsrev}
\bibliography{Fratesi_CVV}

\end{document}